%% file: ms.tex
\documentclass[12pt,preprint2]{aastex}

\usepackage{natbib}

\shorttitle{Rotationally modulated X-rays from accretion}
\shortauthors{Argiroffi et al.}

\begin{document}

\title{The close T Tauri binary system V4046 Sgr: \\
Rotationally modulated X-ray emission from accretion shocks}


\author{C.~Argiroffi\altaffilmark{1,2}, A.~Maggio\altaffilmark{2}, T.~Montmerle\altaffilmark{3}, D.~P.~Huenemoerder\altaffilmark{4},
E.~Alecian\altaffilmark{5}, M.~Audard \altaffilmark{6,7},
J.~Bouvier\altaffilmark{8}, F.~Damiani\altaffilmark{2}, J.-F.~Donati\altaffilmark{9}, S.~G.~Gregory\altaffilmark{10}, M.~G\"udel\altaffilmark{11}, G.~A.~J.~Hussain\altaffilmark{12}, J.~H.~Kastner\altaffilmark{13}, and G.~G.~Sacco\altaffilmark{13}}
\altaffiltext{1}{Dip. di Fisica, Univ. di Palermo, Piazza del Parlamento 1, 90134 Palermo, Italy, email: argi@astropa.unipa.it}
\altaffiltext{2}{INAF - Osservatorio Astronomico di Palermo, Piazza del Parlamento 1, 90134 Palermo, Italy}
\altaffiltext{3}{Institut d'Astrophysique de Paris, 98bis bd Arago, FR 75014 Paris, France}
\altaffiltext{4}{MIT, Kavli Institute for Astrophysics and Space Research, 77 Massachusetts Avenue, Cambridge, MA 02139, USA}
\altaffiltext{5}{Observatoire de Paris, LESIA, 5, place Jules Janssen, F-92195 Meudon Principal Cedex, France}
\altaffiltext{6}{ISDC Data Center for Astrophysics, University of Geneva, Ch. d'Ecogia 16, CH-1290 Versoix, Switzerland}
\altaffiltext{7}{Observatoire de Gen\`eve, University of Geneva, Ch. des Maillettes 51, 1290 Versoix, Switzerland}
\altaffiltext{8}{UJF-Grenoble 1 / CNRS-INSU, Institut de Plan\'etologie et d'Astrophysique de Grenoble (IPAG) UMR 5274, Grenoble, F-38041, France}
\altaffiltext{9}{IRAP-UMR 5277, CNRS \& Univ. de Toulouse, 14 Av. E. Belin, F-31400 Toulouse, France}
\altaffiltext{10}{California Institute of Technology, MC 249-17, Pasadena, CA 91125 USA}
\altaffiltext{11}{University of Vienna, Department of Astronomy, T{\"u}rkenschanzstrasse 17, 1180 Vienna, Austria}
\altaffiltext{12}{ESO, Karl-Schwarzschild-Strasse 2, 85748 Garching bei M\"unchen, Germany}
\altaffiltext{13}{Center for Imaging Science, Rochester Institute of Technology, 54 Lomb Memorial Drive, Rochester, NY 14623, USA}

\begin{abstract}
We report initial results from a quasi-simultaneous X-ray/optical observing campaign targeting V4046~Sgr, a close, synchronous-rotating classical T Tauri star (CTTS) binary in which both components are actively accreting. V4046~Sgr is a strong X-ray source, with the X-rays mainly arising from high-density ($n_{\rm e}\sim10^{11-12}\,{\rm cm^{-3}}$) plasma at temperatures of $3-4$\,MK. Our multiwavelength campaign aims to simultaneously constrain the properties of this X-ray emitting plasma, the large scale magnetic field, and the accretion geometry. In this paper, we present key results obtained via time-resolved X-ray grating spectra, gathered in a 360\,ks {\it XMM-Newton} observation that covered 2.2 system rotations. We find that the emission lines produced by this high-density plasma display periodic flux variations with a measured period, $1.22\pm0.01$\,d, that is precisely half that of the binary star system ($2.42$\,d). The observed rotational modulation can be explained assuming that the high-density plasma occupies small portions of the stellar surfaces, corotating with the stars, and that the high-density plasma is
not azimuthally symmetrically distributed with respect to the rotational axis of each star. These results strongly support models in which high-density, X-ray-emitting CTTS plasma is material heated in accretion shocks, located at the base of accretion flows tied to the system by magnetic field lines.
\end{abstract}
\keywords{Accretion, accretion disks --- Stars: individual: \objectname{V4046 Sgr} --- Stars: magnetic field --- Stars: pre-main sequence --- Stars: variables: T Tauri, Herbig Ae/Be --- X-rays: stars}

\section{Introduction}
\label{intro}

In the context of star formation and evolution, understanding the physics of young low-mass stars is essential. Such stars possess strong magnetic fields that regulate the transfer of mass and angular momentum to and from the circumstellar disk, via accretion and outflow phenomena. Young low-mass stars are also intense sources of high-energy emission (UV and X-rays) that ionizes, heats, and photoevaporates material in the circumstellar disk, thus affecting its physical and chemical evolution and, eventually, the disk lifetime \citep{ErcolanoDrake2008,GortiHollenbach2009}.

Low-mass pre-main sequence stars are classified as classical T~Tauri stars (CTTS) when they still accrete mass from the circumstellar disk. They become weak-line T~Tauri stars (WTTS) when the accretion process ends. Both CTTS and WTTS are bright in X-rays due to the presence of hot coronal plasmas, heated and confined by the intense stellar magnetic fields \citep{FeigelsonMontmerle1999,FavataMicela2003,PreibischKim2005,GudelNaze2009}. It was suggested that in CTTS also the accretion process, beside the coronal magnetic activity, can provide a further X-ray emission mechanism \citep{Ulrich1976,Gullbring1994,Lamzin1999}. Magnetospheric accretion models predict that in CTTS mass transfer from the inner disk onto the star occurs via accretion streams funneled by magnetic flux tubes \citep[e.g.][]{Konigl1991,HartmannHewett1994,BouvierAlencar2007}, where material moves in a almost free fall with typical velocities of $\sim300-500\,{\rm km\,s^{-1}}$. The impact with the stellar atmosphere, usually involving small fractions of the stellar surface, generates shock fronts that heat the infalling material up to temperatures of a few MK, and therefore should yield significant emission in the soft X-ray band ($0.1-1$\,keV). Numerical modeling predicts high $L_{\rm X}$ ($\sim10^{30}\,{\rm erg\,s^{-1}}$) even for low accretion rates ($10^{-10}\,{\rm M_{\odot}\,yr^{-1}}$), indicating that X-ray emission related to the accretion process can rival or exceed coronal emission \citep{GuntherSchmitt2007,SaccoArgiroffi2008} at least in principle.

Strong evidence of accretion-driven X-rays from CTTS has been provided by the observed high densities of the X-ray emitting plasma at $T\sim2-4$\,MK \citep[$n_{\rm e}\sim10^{12}-10^{13}\,{\rm cm^{-3}}$,][]{KastnerHuenemoerder2002,SchmittRobrade2005,GuntherLiefke2006,ArgiroffiMaggio2007,HuenemoerderKastner2007,RobradeSchmitt2007,ArgiroffiFlaccomio2011}. These densities, considering the typical accretion rates and surface filling factors, are compatible with predictions of shock-heated material, and are significantly higher than that of typical quiescent coronal plasmas at temperatures of a few MK \citep[$n_{\rm e}\le10^{10}\,{\rm cm^{-3}}$,][]{NessGudel2004,TestaDrake2004}. Moreover \citet{GudelTelleschi2007} observed a soft X-ray excess in CTTS with respect to WTTS, compatible with the scenario of a further plasma component at a few MK produced by accretion. However other results are discrepant with predictions: the observed $L_{\rm X}$ of the high-density cool-plasma component in CTTS is lower than that predicted from the accretion rate by more than a factor 10 \citep{ArgiroffiMaggio2009,CurranArgiroffi2011}, leaving the coronal component the major contributor to the X-ray emission in CTTS; furthermore in the cool plasma of CTTS the density increases for increasing temperature, at odds with predictions based on a single accretion stream \citep{BrickhouseCranmer2010}. Because of these apparent discrepancies different scenarios were proposed, suggesting that the high-density cool plasma in CTTS could be coronal plasma, confined into magnetic loops, that is somehow modified by the accretion process \citep{GudelSkinner2007,BrickhouseCranmer2010,DupreeBrickhouse2012}.

In addition to containing plasma at a few MK, the shock region is known to be associated with material at $T\sim10^{4}$\,K or more, significantly hotter than the surrounding unperturbed photosphere, as a consequence of the energy locally deposited by the accretion process. This photospheric hot spot produces excess emission in the UV and optical band, which is often rotationally modulated because of the very small filling factor of the accretion-shock region and because accretion streams are usually not symmetric with respect to the rotation axis \citep{BouvierCabrit1993,HerbstHerbst1994,PetrovGahm2001}. Therefore, if the observed high-density X-ray emitting plasma also originates in the accretion shock, then, its X-ray emission might display rotational modulation. Specifically plasma heated in the accretion-shock, observed in the X-rays, could display periodic variations in density, emission measure, average temperatures, absorption, and source optical depth, as a consequence of stellar rotation. First hints of accretion driven X-rays that vary because of the stellar rotation were provided by \citet[][]{ArgiroffiFlaccomio2011} for the star V2129~Oph.

Understanding the origin of this high-density plasma is important, both for constraining the total amount of X-rays emitted in CTTS, and setting the energy balance of the accretion-shock region \citep{SaccoOrlando2010}. Eventually, a definitive confirmation that this plasma component is material heated in the accretion shock, would make its X-ray radiation an insightful tool to probe the physical properties (i.e. density and velocity) of the accretion stream, and to measure the chemical composition of the inner disk material \citep{DrakeTesta2005}.

To search for such X-ray modulation effects we planned and carried out X-ray monitoring of V4046~Sgr, a close binary CTTS system in which both components are actively accreting from a circumbinary disk (see \S~\ref{vsagprop}).

In this work we describe the first results from an {\it XMM-Newton} Large Program (LP) focused on V4046~Sgr, based on time-resolved high-resolution X-ray spectroscopy on timescales down to 1/10 of the system orbital period. To constrain the large-scale magnetic field and the accretion geometry, we also carried out a coordinated multi-wavelength campaign involving photometry, spectroscopy, and spectropolarimetry of V4046~Sgr.

In \S~\ref{project} we summarize the project focused on V4046~Sgr, whose properties are described in \S~\ref{vsagprop}. Details of the data processing and analysis are reported in \S~\ref{obs}. The observing results are presented in \S~\ref{res}, and then discussed in \S~\ref{disc}.

\section{The V4046~Sgr project}
\label{project}

The {\it XMM-Newton} observation of V4046~Sgr consists of a 360\,ks exposure performed on 2009 September 15-19 (Obs-id: 0604860201, 0604860301, and 0604860401). This observation is part of a quasi-simultaneous multi-wavelength campaign (optical photometry with REM/ROSS, 2009 September 1-30; optical spectroscopy with TNG/SARG, 2009 September 10-17; optical spectropolarimetry with CFHT/ESPADONS, 2009 September 2-8), aimed at studying simultaneously the properties of coronal plasmas, stellar magnetic field structure, photospheric spots (both cool spots and hot spots), and the accretion process.

Here we present the results obtained with the {\it XMM-Newton}/RGS specifically aimed at searching for rotational modulation in the accretion-driven X-rays. The results of the entire observing campaign are presented in a series of papers describing, among other results, the properties of the X-ray emitting plasma (A.~Maggio et al., in preparation), maps of the large-scale magnetic field structure and accretion geometry as inferred from optical spectropolarimetry \citep[][ G.~A.~J.~Hussain et al., in preparation, S.~G.~Gregory et al., in preparation]{DonatiGregory2011}, variations in the accretion process over a range of timescales (G.~G.~Sacco et al., in preparation), detection and identification of a distant comoving WTTS system \citep{KastnerSacco2011}. 

\section{V4046~Sgr properties}
\label{vsagprop}

V4046~Sgr is a close CTTS binary system composed of two solar-like mass stars \citep[masses of 0.91 and $0.88\,{\rm M_{\odot}}$, radii of 1.12 and $1.04\,{\rm R_{\odot}}$][]{DonatiGregory2011}, separated by $8.8\,{\rm R_{\odot}}$. The two components are synchronously rotating with a period of 2.42\,d, in circularized orbits \citep{StempelsGahm2004}. V4046~Sgr is estimated to lie at a distance of 73\,pc \citep{TorresQuast2008} and it is viewed with an inclination of $35^{\circ}$ \citep[the angle between the rotation axis and the line of sight,][]{StempelsGahm2004,KastnerZuckerman2008}, with the orbital axis of the binary likely aligned with the individual stellar rotation axes \citep{DonatiGregory2011}. At an age of $\sim10-15$\,Myr \citep{TorresQuast2008,DonatiGregory2011} and classified as a CTTS, V4046~Sgr is still surrounded by a dusty, molecule-rich circumbinary disk \citep{RodriguezKastner2010} from which both the components are actively accreting \citep{StempelsGahm2004}.

A previous {\it Chandra} observation \citep{GuntherLiefke2006} showed that V4046~Sgr has a cool plasma component ($T\approx2-4$\,MK) at high density ($n_{\rm e}\approx0.3-1\times10^{12}\,{\rm cm^{-3}}$), interpreted as material heated in the accretion shock.

At the time of the {\it XMM-Newton} observation, the spectroscopic optical monitoring demonstrated that both components were accreting with a constant rate of $5\times10^{-10}\,{\rm M_{\odot}\,yr^{-1}}$ \citep[inferred from the analysis of the \ion{Ca}{2} IRT, ][]{DonatiGregory2011}. Both components displayed complex magnetic fields \citep[average surface intensity of $\sim200$\,G,][]{DonatiGregory2011}, significantly weaker than that of younger solar-like CTTS \citep[e.g.][]{DonatiSkelly2010}. These magnetic fields are not strong enough to disrupt local disks farther than $1\,R_{\star}$ above stellar surface, thus the formation of circumstellar stellar disks around each component, distinct from the circumbinary disk, may be possible \citep{deValBorroGahm2011}. The accretion process, based on \ion{Ca}{2} IRT, did not show significant rotational modulation, suggesting that post shock material contributing to these lines is symmetrically distributed with respect to stellar poles.

The optical monitoring campaign confirmed the orbital/rotational period ($2.42\,{\rm d}$), and determined the conjunction and quadrature epochs at the time of the {\it XMM-Newton} observation\footnote{The quadrature with primary receding occurred at 2455078.199 HJD}. In this work we adopt the phase reference defined in \citet[][ ${\rm HJD} = 2446998.335+2.4213459\,E$, with phase $0.0$ indicating the quadrature with primary receding]{StempelsGahm2004}. However our optical monitoring revealed a phase shift of 0.069 with respect to that ephemeris, with quadratures occurring at phases 0.93 and 0.43, and conjunctions at phases 0.18 and 0.68 \citep{DonatiGregory2011}.

\section{Observations}
\label{obs}

The {\it XMM-Newton} observation of V4046~Sgr, composed of three observing segments of $\sim120$\,ks each separated by gaps of $\sim50$\,ks, covered 2.2 system rotations. X-ray emitting material heated in the accretion shock is expected to have temperatures of a few MK at most. Therefore to search for X-ray variability possibly produced in the accretion shock we analyzed the {\it XMM-Newton}/RGS spectra, that contain emission lines that specially probe the coolest plasma components.

The RGS spectrograph, composed of two nominally identical gratings (RGS1 and RGS2), covers the $\sim2-38$\,\AA\, wavelength range. The first order spectrum, embracing the $4-38$\,\AA\, band, has a resolution FWHM of 0.06\,\AA, while the second order provides a resolution FWHM of 0.03\,\AA\, in the $\sim2-19$\,\AA\, range. We extracted RGS spectra using the standard {\sc rgsproc} task. Data were filtered discarding time segments affected by high background count rates. The final net exposures of the three observing segments were of 115, 122, 120\,ks, respectively. We then applied the {\sc rgscombine} task to add the RGS1 and RGS2 spectra of the same order. Totally 34800 and 8200 net counts were registered in the first and second order RGS spectra, respectively.

We analyzed the RGS spectra using the IDL package PINTofALE~v2.0 \citep{KashyapDrake2000} and the XSPEC~v12.5 \citep{Arnaud1996} software.
We measured individual line fluxes by fitting simultaneously first and second\footnote{Second order spectrum was used only for lines contained in its wavelength range.} order RGS spectra. Fit procedure was performed in small wavelength intervals ($\Delta \lambda \lesssim 1.0$\,\AA). The adopted best fit function takes into account the RGS line spread function (determined by the matrix response fuction), and the continuum contribution (determined by adding a constant to the line emission, and leaving this constant as a free parameter in the fit).

\section{Results}
\label{res}


The RGS spectra collected during the entire observation (see details in A.~Maggio et al., in preparation) indicate that the main properties of the X-ray emitting plasma of V4046~Sgr are similar to those observed during the previous {\it Chandra} observation \citep{GuntherLiefke2006}: the plasma at $T\sim1-4$\,MK has high density, $n_{\rm e}\sim10^{11}-10^{12}\,{\rm cm^{-3}}$, as determined by the $f/i$ line ratio of He-like triplets of \ion{N}{6}, \ion{O}{7}, and \ion{Ne}{9}\footnote{The measurements of the \ion{Ne}{9} triplet was performed by including in the fit the \ion{Fe}{19} line at 13.52\,\AA, that is anyhow weaker than the \ion{Ne}{9} lines.}.

\input{t1}

\begin{figure*}
\epsscale{2.0}
\plotone{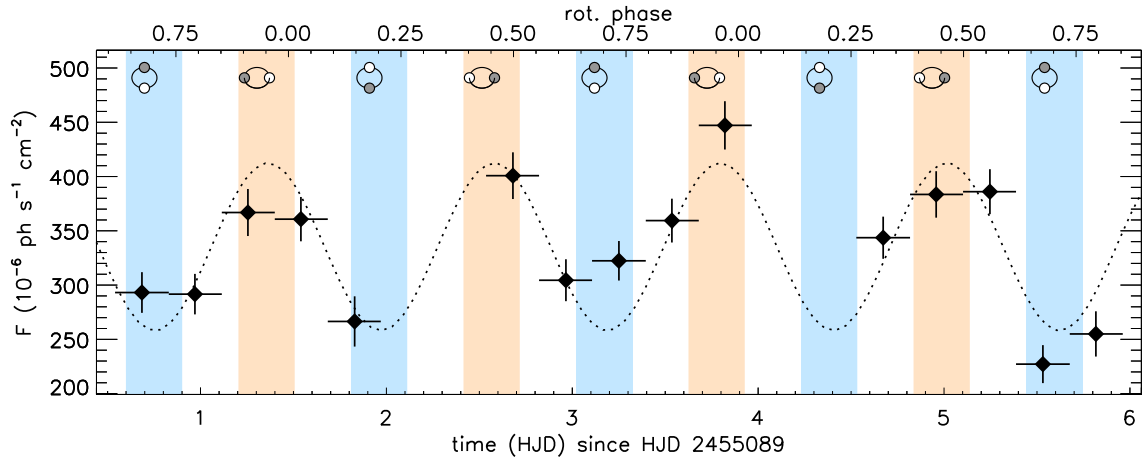}
\caption{Total flux of the cool line set versus time. The set of cool lines is composed of: \ion{Ne}{9} triplet (13.45, 13.55, and 13.70\,\AA), \ion{O}{8} Ly$\alpha$ and Ly$\beta$ (16.00 and 18.98\,\AA), \ion{O}{7} resonance line (21.60\,\AA), and \ion{N}{7} Ly$\alpha$ (24.78\,\AA)). Horizontal error bars represent the time-bin width. Dotted line marks the best-fit sinusoidal function. Orbital/rotational phases are computed according to the ephemeris ${\rm HJD} = 2446998.335+2.4213459\,E$ defined in \citet{StempelsGahm2004}. Vertical dashed lines (dark gray) indicate quadrature and conjunction epochs, with the corresponding schematic views of the system plotted above (white and gray circles represent the primary and secondary components, respectively). Time intervals adopted for extracting spectra corresponding to {\it low} and {\it high} phases are marked by the vertical bands (light blue and light red for the low and high phase, respectively).}
\label{f1}
\end{figure*}

\begin{figure*}
\epsscale{2.0}
\plotone{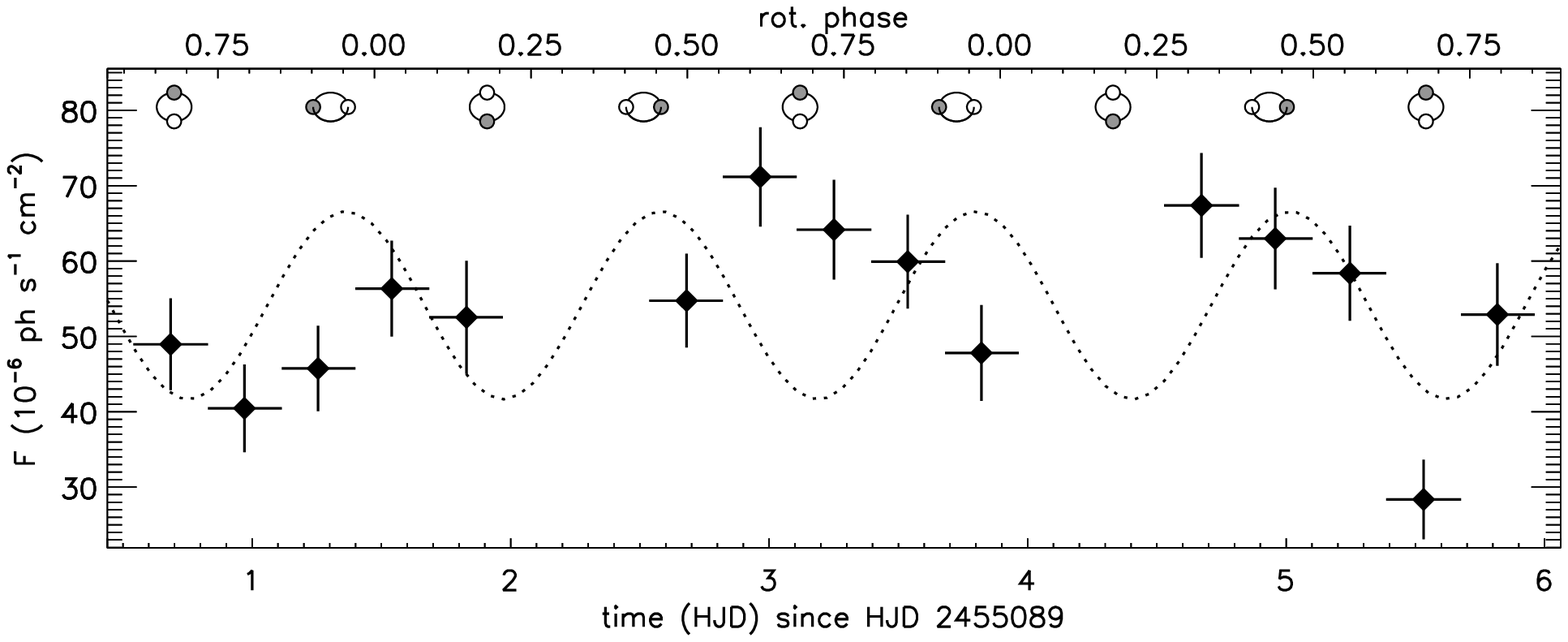}
\caption{Total flux of the hot line set versus time. The set of hot lines is composed of: \ion{Ne}{10} Ly$\alpha$ line at 12.13\,\AA~ and \ion{Fe}{17} line at 15.02\,\AA. Dotted line marks a sinusoidal function with the same period, phase, and relative amplitude obtained from the best fit of the total flux of the cool line set. The hot lines do not show rotational modulation, unlike the cool lines, see Fig.~\ref{f1}, suggesting that their variability is associated with coronal plasma variability.}
\label{f2}
\end{figure*}

\subsection{Time resolved RGS spectra}

To investigate variability on short timescales, we analyzed RGS spectra gathered in time intervals of $\sim25$\,ks (i.e. bins of 0.12 in rotational phase). Totally nine lines have fluxes detected at 1$\sigma$ level in all the time intervals. These lines, and their fluxes at different time intervals, are reported in Table~\ref{t1}. Significant variability on the explored timescales is observed for all the listed lines.

To check for variations in the coolest plasma components we considered lines with peak formation temperature $T_{\rm max}<5$\,MK among the lines reported in Table~\ref{t1}. This sample of lines, named {\it cool} lines, is composed of: the \ion{Ne}{9} triplet (13.45, 13.55, and 13.70\,\AA), \ion{O}{8} Ly$\beta$\footnote{This is blended with an \ion{Fe}{18}, that is however negligible because of the $EMD$ and abundances of the X-ray emitting plasma.} and Ly$\alpha$ (16.00 and 18.98\,\AA), \ion{O}{7} resonance line (21.60\,\AA), and \ion{N}{7} Ly$\alpha$ (24.78\,\AA). Among the lines reported in Table~\ref{t1}, the \ion{Ne}{10} and \ion{Fe}{17} lines stay out of the {\it cool} line sample, because of their $T_{\rm max}$ higher than 5\,MK. Therefore their flux likely includes significant contributions from hot plasma. These two lines compose the {\it hot} line sample.

\input{t2}

To maximize the $S/N$ of the coolest plasma emission we added the measured fluxes of the cool lines for each time interval. This total line flux, plotted in Fig.~\ref{f1}, is variable and the observed modulation is clearly linked to the stellar rotation: the flux is higher near phases 0.0 and 0.5, i.e. quadrature phases, and lower near phases 0.25 and 0.75,  i.e. conjunction phases. To confirm this variability pattern we fitted these observed flux variations with a sinusoid plus a constant. We left all the best-fit function parameters (period, phase, amplitude, and the additive constant) free to vary. We obtained a best-fit period of $1.22\pm0.01$\,d, and an amplitude of $23\pm2\,\%$ with respect to the mean value (Table~\ref{t2}). The inferred period is exactly half the rotational period of the system. As guessed maximum and minimum phases occur approximately at quadrature and conjunction, respectively.
To check whether this observed modulation is effectively linked to the cool plasma emission, and not to a given line emission, we performed the same fit by separately considering the total flux obtained from different and independent cool line subsets. In all the inspected cases (see Table~\ref{t2}) we found the same periodic variability (period, phase, amplitude).

We checked whether this modulation is present also in the emission of hotter plasma by applying the same fit procedure to the total flux of the hot lines, \ion{Ne}{10} and \ion{Fe}{17}. Fit results are reported in Table~\ref{f2}, in this case the periodic modulation is not detected. The observed variability is instead likely dominated by hot (coronal) plasma. Figure~\ref{f2} shows a comparison between \ion{Ne}{10}+\ion{Fe}{17} line variability with modulation observed for the cool lines. The detected X-ray rotational modulation is also not visible in the EPIC lightcurves (A.~Maggio et al., in preparation), even considering only a soft band. The substantial continuum contribution mostly due to the highly variable hot plasma likely masks the rotationally modulated signal. Hence we conclude that the observed X-ray line flux modulation is due to the high-density, cool plasma component.

To understand the nature of the observed variability we searched for variations in the average temperature by considering ratios of lines originating from the same element. All the inspected ratios display significant variability, but are not correlated among themselves, and are not related to the rotational phase. We also searched for variations in the plasma density, probed by the $f/i$ ratio of the \ion{Ne}{9} triplet. This line ratio is approximately constant ($f/i\approx1$, indicating $n_{\rm e}\approx10^{12}\,{\rm cm^{-3}}$) during the entire observation, except for a lower value measured during the third interval of the second segment ($f/i=0.45\pm0.13$, corresponding to $n_{\rm}=(5.2^{+2.0}_{-1.3})\times10^{12}\,{\rm cm^{-3}}$), and a higher value observed during the fourth interval of the third segment ($f/i=3^{+2.5}_{-1.1}$, corresponding to $n_{\rm}<4\times10^{11}\,{\rm cm^{-3}}$). These variations appear to be associated with episodic events, like clumpy accretion flows, and not with a rotational modulation effect.

\input{t3}

\input{t4}

\begin{figure*}
\epsscale{2.0}
\plotone{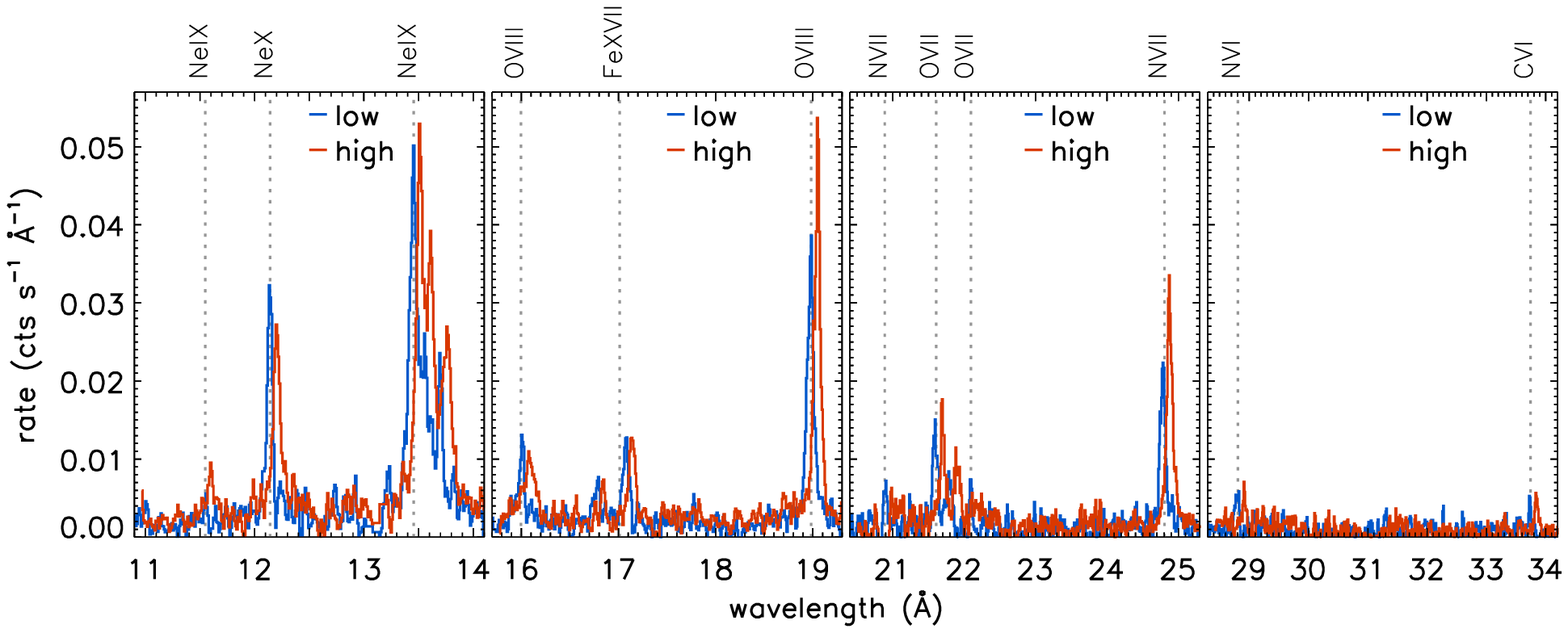}
\caption{RGS spectra corresponding to minimum (low) and maximum (high) phases, corresponding to exposure times of 84 and 94\,ks respectively. For clarity reasons the two spectra are slightly smoothed, and the spectrum of the high-flux phase is shifted toward longer wavelengths by $0.1$\,\AA.}
\label{f3}
\end{figure*}

\subsection{RGS spectra at different phases}

The total flux of the cool lines from V4046~Sgr displayed variations in time linked to the stellar rotation. To investigate the differences of the X-ray emitting plasma corresponding to epochs of low and high fluxes of the cool lines, we added RGS data collected at the same phases with respect to the X-ray rotational modulation. We extracted two RGS spectra obtained by adding all the events registered during time intervals centered on maximum and minimum times, with duration of one fourth of the observed X-ray period (integration time intervals are shown in Fig.~\ref{f1}). The two resulting {\it low} and {\it high} spectra, whose exposure times are 84 and 94\,ks respectively, are shown in Fig.~\ref{f3}, while the measured line fluxes, detected at 1$\sigma$ level in the two spectra, are listed in Table~\ref{t3}.

\begin{figure}
\epsscale{1.0}
\plotone{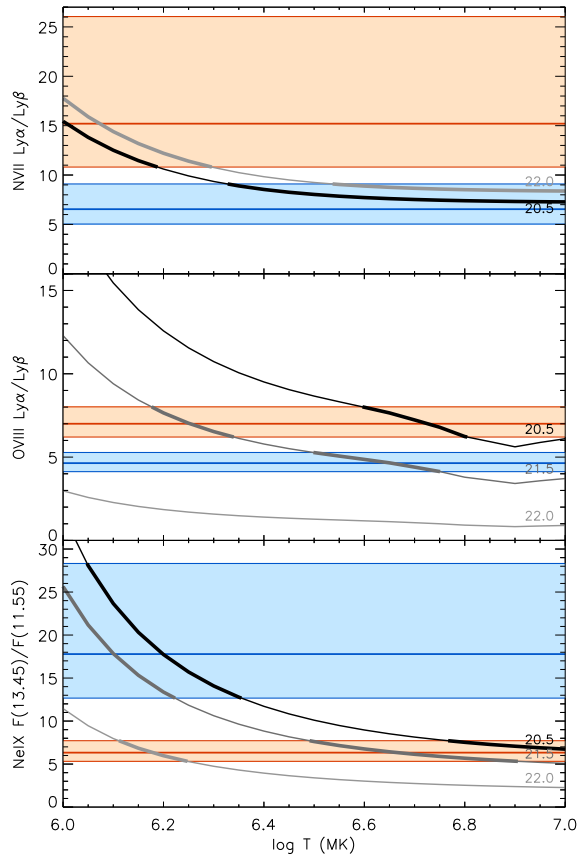}
\caption{Ly$\alpha$ to Ly$\beta$ photon flux ratio for the \ion{N}{7} and \ion{O}{8} H-like ions and $13.45$\,\AA~ to $11.55$\,\AA~ for the \ion{Ne}{9} He-like ion, vs plasma temperature. Horizontal bands indicate values measured during {\it high} (red) and {\it low} (blue) phases. Black, dark gray, and light gray lines represent predicted values for different absorptions $N_{\rm H}$, with labels reporting the corresponding $\log N_{\rm H}$ value, and with thicker lines marking curve portions compatible with the observed ratios.}
\label{f4}
\end{figure}

We searched for differences in the {\it low} and {\it high} spectra to investigate how the emitting plasma properties vary between these two phases. The two spectra display significantly different photon flux ratios of \ion{N}{7}, \ion{O}{8}, \ion{Ne}{9} lines, as reported in Table~\ref{t4}. In principle, these line ratios may vary due to changes of absorption, plasma temperature, or source optical depth. In Fig.~\ref{f4} we plot the measured line ratios together with the values predicted in the optically thin regime for different temperatures and different hydrogen column densities, $N_{\rm H}$.

Absorption can change line ratios because, on average, lines at longer wavelengths suffer larger attenuation for increasing $N_{\rm H}$. The two \ion{N}{7} lines considered here are an exception, because the absorption cross section of the interstellar medium has the oxygen K-shell edge \citep[23.3\,\AA, e.g.][]{WilmsAllen2000ApJ} located between their wavelengths, making the longer wavelength line, the Ly$\alpha$ (24.78\,\AA), slightly less absorbed than the Ly$\beta$ (20.91\,\AA). The two lines however suffer very similar absorption, making the absorption effect of little relevance in the case of the \ion{N}{7} Ly$\alpha$/Ly$\beta$ ratio (as can be seen from the upper panel of fig.~\ref{f4}, where the curves predicted for different $N_{\rm H}$ are very similar). Therefore any change in this line ratio, as that observed, is hardly explained in terms of $N_{\rm H}$ variability. Instead, an $N_{\rm H}$ decrease from the {\it low} to the {\it high} state might explain the variation of the \ion{O}{8} line ratios, but an opposite $N_{\rm H}$ variation should be invoked to justify the \ion{Ne}{9} variability (middle and lower panels of fig.~\ref{f4}). All this findings indicate that the hydrogen column density toward the source appears to be unchanged, and that line ratio variability is produced by a different mechanism. This conclusion is supported by the similar fluxes between {\it low} and {\it high} spectra measured for the two lines at long wavelengths (\ion{N}{6} at 28.8\,\AA\, and \ion{C}{6} at 33.7\,\AA), the most affected by absorption, and it is also confirmed by the full fledged analysis of the EPIC data presented in A.~Maggio et al., in preparation, where $N_{\rm H}$ is found to vary by only a factor 2 over the whole observation around a mean value of  $3\times10^{20}\,{\rm cm^{-2}}$ (i.e. $\log N_{\rm H} = 20.5$).

The three explored ratios depend also on temperature, because of the different energy of the upper levels of the two electronic transitions considered in each ratio. In this respect, the three ratios do not vary consistently. In fact, a temperature decrease from the {\it low} state to the {\it high} state could explain the increasing \ion{N}{7} and \ion{O}{8} line ratios, but not the variation \ion{Ne}{9} lines. The derivation of the plasma model (see A.~Maggio et al., in preparation) is beyond the scope of this work, but we anticipate here that the $EMD$ does not appear to vary enough between the two phases to justify the observed variations of the line ratios. Moreover the average plasma temperature ($\log T \sim 6.6-6.7$), together with the measured $N_{\rm H}$, indicates that, in some phases, line ratios are not compatible with the optically thin limit, irrespective of the nature of their variability.

Optical depth effects can change line ratios because each line optical depth is directly proportional to the oscillator strength of the transition \citep{Acton1978}. Therefore, if optically thin emission does not apply, transitions with very different oscillator strength
may suffer different attenuation/enhancement, with stronger effects occurring in lines with higher oscillator strengths. In the inspected ratios the lines with higher oscillator strength are the $Ly\alpha$ line of \ion{N}{7} and \ion{O}{8}, and the $13.45$\,\AA~line of \ion{Ne}{9} \citep[e.g.][]{TestaDrake2007}. Non negligible optical depth is expected in strong X-ray resonance lines produced from shock-heated plasma in CTTS \citep{ArgiroffiMaggio2009}. The observed variable ratios might indicate that some lines are affected by a changing optical depth. Since the expected attenuation/enhancement with respect to optically thin emission depends on the source geometry and viewing angle, stellar rotation can produce periodic changes in the line opacity, and hence in the observed ratios. However, once again the three ratios do not vary in the same direction, with \ion{N}{7} and \ion{O}{8} ratios being higher in the {\it low} state, whereas the ratio of the slightly hotter \ion{Ne}{9} lines results higher in the {\it high} phase.

Summarizing we stress the significant variations observed in line ratios between {\it low} and {\it high} phases. The origin of these variations remains unclear. If these variations were due to changes in plasma temperature or absorption, a coherent behavior would be expected for the three ratios, and this is not the case. Opacity effects instead can operate in a more complex way, provoking both line enhancements or reductions, depending on the source geometry and viewing angle. This hypothesis is therefore the most intriguing, especially considering that
in some phases line ratios are discrepant from the value expected in the optically-thin limit.

\section{Discussion}
\label{disc}

The main result of the time resolved spectral analysis of the X-ray emitting plasma from V4046~Sgr (\S~\ref{res}) is that the high-density plasma component at $3-4$\,MK is rotationally modulated with a period of half the system orbital period, with maximum and minimum phases occurring at quadrature and conjunction epochs, respectively. The observed X-ray rotational modulation indicates that this high-density plasma component is not symmetrically distributed with respect to the stellar rotational axes. We also found that strong emission lines from this plasma component provide some indications of non negligible optical depth effects, and that the periodic modulation appears to be associated with variations in the source optical depth, as evidenced by the significant variations in line ratios sensitive to optical depth observed between {\it low} and {\it high} phases.

The strongest X-ray emission lines, produced by shock-heated material in CTTS, are expected to have non-negligible optical depth due to the high density and typical size of the post-shock region \citep{ArgiroffiMaggio2009}. Moreover the optical depth should vary if the viewing geometry of the post shock region changes.

Hints of non-negligible optical depth observed in the strongest X-ray lines of V4046~Sgr indicate that the high-density plasma is mostly concentrated in a compact portion of the stellar surface, as predicted for the post-shock material. Moreover, the variability of the optical depth can be naturally explained with the changing viewing geometry of the volume occupied by the high-density plasma during stellar rotation. This scenario requires plasma confinement by the stellar magnetic fields.

We observed an X-ray period of half the system orbital period, as already observed for accretion indicators \citep[e.g.][]{VrbaChugainov1993,KurosawaHarries2005} and X-ray emission \citep{FlaccomioMicela2005} from some CTTS. That could be explained, in the case of V4046~Sgr, by different scenarios. If the X-ray emitting plasma is located only in one of the two system components, then a period of half the rotational period is observed when there are two accretion-shock regions on the stellar surface at opposite longitudes, or there is only one accretion-shock region and the maximum X-ray flux is observed when the base of the accretion stream is viewed sideways \citep[][ a configuration that occurs twice in one stellar rotation]{ArgiroffiFlaccomio2011}.

Considering the system symmetry and the accretion geometry previously suggested by \citet{StempelsGahm2004}, it is conceivable that both components possess similar amounts of high-density cool plasma. In this scenario the half period can be naturally explained assuming that the location on each stellar surface of this plasma, compact and not azimuthally symmetric with respect to each stellar rotation axis, is symmetric for $180^{\circ}$ rotations with respect to the binary rotation axis.

The simultaneous optical monitoring campaign indicated that the two components have similar accretion rates, validating the assumption that the two components possess similar amounts of high-density plasma. However, the optical accretion spots, probed by \ion{Ca}{2} IRT, did not show rotational modulation \citep{DonatiGregory2011}. Therefore accretion regions emitting \ion{Ca}{2} should be symmetrically distributed with respect to the stellar poles. This scenario, different from that obtained from the X-ray data, could be reconciled considering that X-rays are likely produced only by a fraction of the entire accretion-shock region \citep{SaccoOrlando2010}.

In conclusion, our {\it XMM-Newton}/RGS data of the V4046~Sgr close binary system have shown for the first time the rotational modulation of X-ray lines characteristic of a cool, high-density plasma corotating with the stars. This strongly support the accretion-driven X-ray emission scenario, in which the high-density cool plasma of CTTS is material heated in the accretion shock. It moreover suggests that the accretion flow is channeled by magnetic field lines anchored on the stars, along small magnetic tubes. This is consistent with the general framework of magnetic accretion, but brings new insights into the accretion mechanism in close binary systems of CTTS.


\acknowledgments
This work is based on observations obtained with XMM-Newton, an ESA science mission with instruments and contributions directly funded by ESA Member States and NASA. C.A., A.M., and F.D. acknowledge financial contribution from the agreement ASI-INAF I/009/10/0.

\end{document}

%% file: t1.tex
\begin{deluxetable}{c@{\hspace{0mm}}c@{\hspace{0mm}}r@{$\pm$}l@{\hspace{1mm}}r@{$\pm$}l@{\hspace{1mm}}r@{$\pm$}l@{\hspace{1mm}}r@{$\pm$}l@{\hspace{1mm}}r@{$\pm$}l@{\hspace{1mm}}r@{$\pm$}l@{\hspace{1mm}}r@{$\pm$}l@{\hspace{1mm}}r@{$\pm$}l@{\hspace{1mm}}r@{$\pm$}l@{\hspace{1mm}}}
\tabletypesize{\scriptsize \label{t1}}
\tablewidth{0pt}
\tablecaption{Observed line fluxes at different time intervals}
\tablehead{\colhead{} & \colhead{} & \multicolumn{2}{c}{ Ne~X } & \multicolumn{2}{c}{ Ne~IX } & \multicolumn{2}{c}{ Ne~IX } & \multicolumn{2}{c}{ Ne~IX } & \multicolumn{2}{c}{ Fe~XVII } & \multicolumn{2}{c}{ O~VIII } & \multicolumn{2}{c}{ O~VIII } & \multicolumn{2}{c}{ O~VII } & \multicolumn{2}{c}{ N~VII } \\
\colhead{} & \colhead{rot.} & \multicolumn{2}{c}{ 12.13\,\AA} & \multicolumn{2}{c}{ 13.45\,\AA} & \multicolumn{2}{c}{ 13.55\,\AA} & \multicolumn{2}{c}{ 13.70\,\AA} & \multicolumn{2}{c}{ 15.02\,\AA} & \multicolumn{2}{c}{ 16.01\,\AA} & \multicolumn{2}{c}{ 18.97\,\AA} & \multicolumn{2}{c}{ 21.60\,\AA} & \multicolumn{2}{c}{ 24.78\,\AA} \\
\colhead{} & \colhead{phase} & \multicolumn{2}{c}{ 6.3\,MK} & \multicolumn{2}{c}{ 4.0\,MK} & \multicolumn{2}{c}{ 3.5\,MK} & \multicolumn{2}{c}{ 4.0\,MK} & \multicolumn{2}{c}{ 5.6\,MK} & \multicolumn{2}{c}{ 3.2\,MK} & \multicolumn{2}{c}{ 3.2\,MK} & \multicolumn{2}{c}{ 2.0\,MK} & \multicolumn{2}{c}{ 2.0\,MK}}
\startdata
$F_{\rm int1/seg1}$ & 0.67 &   34.0 &    5.2 &   58.3 &    6.6 &   21.1 &    5.2 &   21.3 &    4.8 &   14.9 &    3.2 &   15.3 &    3.3 &   90.6 &    8.9 &   18.9 &    8.9 &   67.6 &    9.4 \\
$F_{\rm int2/seg1}$ & 0.79 &   33.0 &    5.0 &   76.0 &    7.0 &   17.7 &    5.1 &   16.8 &    4.3 &    7.4 &    3.0 &   12.8 &    3.7 &   81.8 &    8.1 &   17.0 &    9.3 &   69.5 &    9.4 \\
$F_{\rm int3/seg1}$ & 0.91 &   31.7 &    4.7 &   56.1 &    6.9 &   34.3 &    6.2 &   25.3 &    5.5 &   14.0 &    3.2 &   17.6 &    3.7 &  101.0 &    9.1 &   42.5 &   11.7 &   89.9 &   11.0 \\
$F_{\rm int4/seg1}$ & 0.03 &   45.9 &    5.4 &   65.7 &    6.9 &   31.6 &    5.8 &   24.9 &    4.9 &   10.5 &    3.3 &    9.8 &    3.3 &  118.2 &    9.7 &   48.3 &   11.0 &   62.3 &    9.2 \\
$F_{\rm int5/seg1}$ & 0.15 &   40.9 &    6.6 &   59.8 &    8.3 &   22.0 &    6.8 &   27.1 &    6.3 &   11.6 &    3.6 &   22.3 &    5.0 &   52.4 &    8.9 &   30.6 &   12.7 &   52.3 &   10.8 \\
$F_{\rm int1/seg2}$ & 0.50 &   40.2 &    5.2 &   71.0 &    7.4 &   33.8 &    6.1 &   37.3 &    5.6 &   14.5 &    3.4 &   10.9 &    3.7 &   95.6 &    8.4 &   33.2 &   11.8 &  119.1 &   10.7 \\
$F_{\rm int2/seg2}$ & 0.62 &   54.3 &    5.5 &   37.7 &    6.1 &   30.3 &    5.7 &   35.9 &    5.3 &   16.9 &    3.6 &   13.1 &    3.7 &   92.0 &    8.5 &   31.2 &   10.1 &   64.3 &    9.3 \\
$F_{\rm int3/seg2}$ & 0.73 &   45.6 &    5.5 &   43.1 &    6.3 &   45.4 &    6.2 &   20.2 &    4.9 &   18.5 &    3.7 &   20.4 &    4.0 &  101.9 &    8.4 &   19.0 &    7.2 &   72.5 &    9.5 \\
$F_{\rm int4/seg2}$ & 0.85 &   42.5 &    5.2 &   55.4 &    6.6 &   31.7 &    6.0 &   29.5 &    5.3 &   17.4 &    3.4 &   10.6 &    3.4 &  112.6 &    9.4 &   42.6 &    9.6 &   77.0 &   10.4 \\
$F_{\rm int5/seg2}$ & 0.97 &   42.6 &    5.7 &   75.1 &    8.2 &   60.8 &    7.5 &   39.7 &    6.1 &    5.2 &    2.8 &   18.8 &    3.7 &  121.3 &    9.6 &   24.1 &    9.7 &  107.4 &   11.6 \\
$F_{\rm int1/seg3}$ & 0.32 &   61.6 &    6.0 &   46.3 &    6.0 &   26.6 &    5.4 &   20.7 &    4.6 &    5.8 &    3.4 &    9.7 &    3.7 &  135.8 &    9.8 &   35.0 &    9.1 &   69.4 &   10.0 \\
$F_{\rm int2/seg3}$ & 0.44 &   48.3 &    5.6 &   68.5 &    7.2 &   41.3 &    6.2 &   32.5 &    5.1 &   14.7 &    3.7 &   12.8 &    3.7 &   98.3 &    8.7 &   53.3 &   12.3 &   76.7 &   10.0 \\
$F_{\rm int3/seg3}$ & 0.56 &   43.3 &    5.3 &   66.6 &    7.1 &   20.4 &    5.3 &   35.9 &    5.1 &   15.1 &    3.4 &   16.6 &    3.9 &  112.0 &    9.1 &   39.9 &   10.4 &   94.6 &   10.9 \\
$F_{\rm int4/seg3}$ & 0.68 &   23.7 &    4.3 &   36.9 &    5.9 &    9.9 &    4.3 &   30.7 &    5.1 &    4.7 &    3.0 &   11.9 &    3.1 &   69.6 &    7.7 &   24.9 &    9.9 &   43.4 &    7.8 \\
$F_{\rm int5/seg3}$ & 0.79 &   39.1 &    5.8 &   41.9 &    6.4 &   21.4 &    6.1 &   20.7 &    4.9 &   13.8 &    3.5 &   13.7 &    3.7 &   75.0 &    8.9 &   26.0 &   11.9 &   56.4 &   10.0 \\
\enddata
\tablecomments{~For each line, the Table head reports: ion, wavelength, and maximum formation temperature. $F_{\rm int {\it i}/seg {\it j}}$ refers to the line fluxes measured in the $i-$th interval of the $j-$th observing segment. For each time interval, the listed phase corresponds to the central time of the bin. Line fluxes are in $10^{-6}\,{\rm ph\,s^{-1}\,cm^{-2}}$. Errors correspond to 1$\sigma$.}
\end{deluxetable}

%% file: t2.tex
\begin{deluxetable}{ccccccc}
\tabletypesize{\footnotesize \label{t2}}
\tablewidth{0pt}
\tablecaption{Best fit parameters}
\tablehead{
\colhead{Set name} & \colhead{Line set}                         & \colhead{P\tablenotemark{a} (d)} & \colhead{A\tablenotemark{b} (\%)} & \colhead{$t_{\rm max}$\tablenotemark{c} (d)} & \colhead{($\chi^2_{\rm red,1}$)\tablenotemark{d}} & \colhead{($\chi^2_{\rm red,2}$)\tablenotemark{e}}
}
\startdata
cool lines       & \ion{Ne}{9}+\ion{O}{8}+\ion{O}{7}+\ion{N}{7} & $1.22\pm0.01$  & $23\pm2$  & $1.36\pm0.05$ & 2.91 & 8.93 \\
cool line subset & \ion{O}{8}+\ion{O}{7}+\ion{N}{7}             & $1.21\pm0.02$  & $22\pm3$  & $1.35\pm0.06$ & 2.31 & 5.50 \\
cool line subset & \ion{O}{7}+\ion{N}{7}                        & $1.22\pm0.03$  & $30\pm5$  & $1.30\pm0.05$ & 0.65 & 2.89 \\
cool line subset & \ion{Ne}{9}                                  & $1.22\pm0.02$  & $25\pm4$  & $1.40\pm0.06$ & 2.33 & 5.12 \\
\hline
hot lines        & \ion{Ne}{10}+\ion{Fe}{17}                    & $1.12\pm0.03$  & $14\pm5$  & $1.61\pm0.10$ & 3.63 & 3.46 \\
\enddata
\tablenotetext{a}{Period.}
\tablenotetext{b}{Amplitude.}
\tablenotetext{c}{First epoch of maximum flux after observation start, since HJD 2455089.}
\tablenotetext{d}{Reduced $\chi^2$ obtained with a sinusoid as best-fit function (4 free parameters).}
\tablenotetext{e}{Reduced $\chi^2$ obtained with a constant as best-fit function (1 free parameters).}
\end{deluxetable}

%% file: t3.tex
\begin{deluxetable}{lccr@{$\pm$}lr@{$\pm$}l}
\tabletypesize{\scriptsize \label{t3}}
\tablewidth{0pt}
\tablecaption{Observed line fluxes at different phases}
\tablehead{\multicolumn{1}{c}{Ion} & \multicolumn{1}{c}{$\lambda$} & \multicolumn{1}{c}{$T_{\rm max}$} & \multicolumn{2}{c}{$F_{Low}$} & \multicolumn{2}{c}{$F_{High}$} \\
 & \multicolumn{1}{c}{(\AA)} & \multicolumn{1}{c}{(MK)} & \multicolumn{2}{c}{$({\rm 10^{-6}\,ph\,s^{-1}\,cm^{-2}})$} & \multicolumn{2}{c}{$({\rm 10^{-6}\,ph\,s^{-1}\,cm^{-2}})$}}
\startdata
 Ne~X  &  10.24 &  6.3 &    5.9 &    1.3 &    4.1 &    1.2 \\
 Ne~IX  &  11.55 &  4.0 &    4.1 &    1.6 &   11.9 &    2.1 \\
 Ne~X  &  12.13 &  6.3 &   46.2 &    3.0 &   50.6 &    3.0 \\
 Fe~XXI  &  12.28 & 10.0 &    4.3 &    1.6 &    3.7 &    1.8 \\
 Ne~IX  &  13.45 &  4.0 &   72.2 &    4.0 &   75.3 &    3.9 \\
 Fe~XIX  &  13.52 & 10.0 &   16.3 &    3.3 &    9.4 &    3.3 \\
 Ne~IX  &  13.55 &  3.5 &   27.3 &    3.1 &   49.4 &    3.5 \\
 Ne~IX  &  13.70 &  4.0 &   29.9 &    2.9 &   41.7 &    3.1 \\
 Fe~XVIII  &  14.20 &  7.9 &    5.1 &    1.6 &    3.8 &    1.5 \\
 Fe~XVII  &  15.02 &  5.6 &   13.6 &    1.8 &   14.3 &    1.9 \\
 O~VIII  &  15.18 &  3.2 &    4.3 &    1.7 &    7.9 &    1.8 \\
 O~VIII  &  16.01 &  3.2 &   19.5 &    2.1 &   16.2 &    2.0 \\
 Fe~XVII  &  16.78 &  5.0 &   11.4 &    2.1 &   10.3 &    1.9 \\
 Fe~XVII  &  17.05 &  5.0 &   12.7 &    2.4 &   14.1 &    2.4 \\
 Fe~XVII  &  17.10 &  5.0 &    5.8 &    2.6 &    8.9 &    2.6 \\
 O~VIII  &  18.97 &  3.2 &   90.5 &    4.7 &  113.9 &    4.8 \\
 N~VII  &  20.91 &  2.2 &   11.0 &    3.1 &    7.0 &    2.9 \\
 O~VII  &  21.60 &  2.0 &   24.8 &    4.0 &   46.0 &    6.1 \\
 O~VII  &  21.81 &  2.0 &   17.3 &    5.9 &   24.4 &    6.3 \\
 O~VII  &  22.10 &  2.0 &   12.2 &    3.9 &   10.3 &    3.3 \\
 N~VII  &  24.78 &  2.0 &   72.0 &    5.4 &  106.6 &    6.0 \\
 N~VI  &  28.79 &  1.6 &   25.0 &    4.5 &   17.9 &    4.5 \\
 N~VI  &  29.08 &  1.3 &   12.7 &    4.1 &    7.5 &    3.9 \\
 C~VI  &  33.73 &  1.6 &   25.3 &    5.2 &   35.6 &    5.7 \\
\enddata
\tablecomments{~Line fluxes measured during the {\it high} and {\it low} phases are listed. Flux errors correspond to 1$\sigma$.}
\end{deluxetable}

%% file: t4.tex
\begin{deluxetable}{crr}
\tabletypesize{\normalsize \label{t4}}
\tablewidth{0pt}
\tablecaption{Line flux ratios at different phases}
\tablehead{\multicolumn{1}{c}{Line flux ratio} & \multicolumn{1}{c}{$R_{Low}$} & \multicolumn{1}{c}{$R_{High}$}}
\startdata
\ion{Ne}{9} (13.45\,\AA) / \ion{Ne}{9} (11.55\,\AA)                   & $18^{+11}_{-5}$     & $6.4^{+1.4}_{-1.0}$ \\
\ion{O}{8} L$\alpha$ (18.97\,\AA) / \ion{O}{8} Ly$\beta$ (16.01\,\AA) & $4.6^{+0.6}_{-0.5}$ & $7.0^{+1.1}_{-0.8}$  \\
\ion{N}{7} L$\alpha$ (24.78\,\AA) / \ion{N}{7} Ly$\beta$ (20.91\,\AA) & $6.5^{+2.6}_{-1.5}$ & $15^{+11}_{-5}$ \\
\enddata
\tablecomments{~Ratio errors correspond to 68\% confidence level.}
\end{deluxetable}

%% file: ms.bbl
\begin{thebibliography}{50}
\expandafter\ifx\csname natexlab\endcsname\relax\def\natexlab#1{#1}\fi

\bibitem[{{Acton}(1978)}]{Acton1978}
{Acton}, L.~W. 1978, \apj, 225, 1069

\bibitem[{{Argiroffi} {et~al.}(2011){Argiroffi}, {Flaccomio}, {Bouvier},
  {Donati}, {Getman}, {Gregory}, {Hussain}, {Jardine}, {Skelly}, \&
  {Walter}}]{ArgiroffiFlaccomio2011}
{Argiroffi}, C., {Flaccomio}, E., {Bouvier}, J., {Donati}, J.-F., {Getman},
  K.~V., {Gregory}, S.~G., {Hussain}, G.~A.~J., {Jardine}, M.~M., {Skelly},
  M.~B., \& {Walter}, F.~M. 2011, \aap, 530, A1

\bibitem[{{Argiroffi} {et~al.}(2007){Argiroffi}, {Maggio}, \&
  {Peres}}]{ArgiroffiMaggio2007}
{Argiroffi}, C., {Maggio}, A., \& {Peres}, G. 2007, \aap, 465, L5

\bibitem[{{Argiroffi} {et~al.}(2009){Argiroffi}, {Maggio}, {Peres}, {Drake},
  {L{\'o}pez-Santiago}, {Sciortino}, \& {Stelzer}}]{ArgiroffiMaggio2009}
{Argiroffi}, C., {Maggio}, A., {Peres}, G., {Drake}, J.~J.,
  {L{\'o}pez-Santiago}, J., {Sciortino}, S., \& {Stelzer}, B. 2009, \aap, 507,
  939

\bibitem[{{Arnaud}(1996)}]{Arnaud1996}
{Arnaud}, K.~A. 1996, in Astronomical Society of the Pacific Conference Series,
  Vol. 101, Astronomical Data Analysis Software and Systems V, ed.
  {G.~H.~Jacoby \& J.~Barnes}, 17

\bibitem[{{Bouvier} {et~al.}(2007){Bouvier}, {Alencar}, {Harries},
  {Johns-Krull}, \& {Romanova}}]{BouvierAlencar2007}
{Bouvier}, J., {Alencar}, S.~H.~P., {Harries}, T.~J., {Johns-Krull}, C.~M., \&
  {Romanova}, M.~M. 2007, Protostars and Planets V, 479

\bibitem[{{Bouvier} {et~al.}(1993){Bouvier}, {Cabrit}, {Fernandez}, {Martin},
  \& {Matthews}}]{BouvierCabrit1993}
{Bouvier}, J., {Cabrit}, S., {Fernandez}, M., {Martin}, E.~L., \& {Matthews},
  J.~M. 1993, \aap, 272, 176

\bibitem[{{Brickhouse} {et~al.}(2010){Brickhouse}, {Cranmer}, {Dupree}, {Luna},
  \& {Wolk}}]{BrickhouseCranmer2010}
{Brickhouse}, N.~S., {Cranmer}, S.~R., {Dupree}, A.~K., {Luna}, G.~J.~M., \&
  {Wolk}, S. 2010, \apj, 710, 1835

\bibitem[{{Curran} {et~al.}(2011){Curran}, {Argiroffi}, {Sacco}, {Orlando},
  {Peres}, {Reale}, \& {Maggio}}]{CurranArgiroffi2011}
{Curran}, R.~L., {Argiroffi}, C., {Sacco}, G.~G., {Orlando}, S., {Peres}, G.,
  {Reale}, F., \& {Maggio}, A. 2011, \aap, 526, A104

\bibitem[{{de Val-Borro} {et~al.}(2011){de Val-Borro}, {Gahm}, {Stempels}, \&
  {Pepli{\'n}ski}}]{deValBorroGahm2011}
{de Val-Borro}, M., {Gahm}, G.~F., {Stempels}, H.~C., \& {Pepli{\'n}ski}, A.
  2011, \mnras, 372

\bibitem[{{Donati} {et~al.}(2011){Donati}, {Gregory}, {Montmerle}, {Maggio},
  {Argiroffi}, {Sacco}, {Hussain}, {Kastner}, {Alencar}, {Audard}, {Bouvier},
  {Damiani}, {G{\"u}del}, {Huenemoerder}, \& {Wade}}]{DonatiGregory2011}
{Donati}, J.-F., {Gregory}, S.~G., {Montmerle}, T., {Maggio}, A., {Argiroffi},
  C., {Sacco}, G., {Hussain}, G., {Kastner}, J., {Alencar}, S.~H.~P., {Audard},
  M., {Bouvier}, J., {Damiani}, F., {G{\"u}del}, M., {Huenemoerder}, D., \&
  {Wade}, G.~A. 2011, \mnras, 417, 1747

\bibitem[{{Donati} {et~al.}(2010){Donati}, {Skelly}, {Bouvier}, {Gregory},
  {Grankin}, {Jardine}, {Hussain}, {M{\'e}nard}, {Dougados}, {Unruh},
  {Mohanty}, {Auri{\`e}re}, {Morin}, \& {Far{\`e}s}}]{DonatiSkelly2010}
{Donati}, J.-F., {Skelly}, M.~B., {Bouvier}, J., {Gregory}, S.~G., {Grankin},
  K.~N., {Jardine}, M.~M., {Hussain}, G.~A.~J., {M{\'e}nard}, F., {Dougados},
  C., {Unruh}, Y., {Mohanty}, S., {Auri{\`e}re}, M., {Morin}, J., \&
  {Far{\`e}s}, R. 2010, \mnras, 409, 1347

\bibitem[{{Drake} {et~al.}(2005){Drake}, {Testa}, \&
  {Hartmann}}]{DrakeTesta2005}
{Drake}, J.~J., {Testa}, P., \& {Hartmann}, L. 2005, \apjl, 627, L149

\bibitem[{{Dupree} {et~al.}(2012){Dupree}, {Brickhouse}, {Cranmer}, {Luna},
  {Schneider}, {Bessell}, {Bonanos}, {Crause}, {Lawson}, {Mallik}, \&
  {Schuler}}]{DupreeBrickhouse2012}
{Dupree}, A.~K., {Brickhouse}, N.~S., {Cranmer}, S.~R., {Luna}, G.~J.~M.,
  {Schneider}, E.~E., {Bessell}, M.~S., {Bonanos}, A., {Crause}, L.~A.,
  {Lawson}, W.~A., {Mallik}, S.~V., \& {Schuler}, S.~C. 2012, \apj, in press

\bibitem[{{Ercolano} {et~al.}(2008){Ercolano}, {Drake}, {Raymond}, \&
  {Clarke}}]{ErcolanoDrake2008}
{Ercolano}, B., {Drake}, J.~J., {Raymond}, J.~C., \& {Clarke}, C.~C. 2008,
  \apj, 688, 398

\bibitem[{{Favata} \& {Micela}(2003)}]{FavataMicela2003}
{Favata}, F., \& {Micela}, G. 2003, \ssr, 108, 577

\bibitem[{{Feigelson} \& {Montmerle}(1999)}]{FeigelsonMontmerle1999}
{Feigelson}, E.~D., \& {Montmerle}, T. 1999, \araa, 37, 363

\bibitem[{{Flaccomio} {et~al.}(2005){Flaccomio}, {Micela}, {Sciortino},
  {Feigelson}, {Herbst}, {Favata}, {Harnden}, \&
  {Vrtilek}}]{FlaccomioMicela2005}
{Flaccomio}, E., {Micela}, G., {Sciortino}, S., {Feigelson}, E.~D., {Herbst},
  W., {Favata}, F., {Harnden}, Jr., F.~R., \& {Vrtilek}, S.~D. 2005, \apjs,
  160, 450

\bibitem[{{Gorti} \& {Hollenbach}(2009)}]{GortiHollenbach2009}
{Gorti}, U., \& {Hollenbach}, D. 2009, \apj, 690, 1539

\bibitem[{{G{\"u}del} \& {Naz{\'e}}(2009)}]{GudelNaze2009}
{G{\"u}del}, M., \& {Naz{\'e}}, Y. 2009, \aapr, 17, 309

\bibitem[{{G{\"u}del} {et~al.}(2007){G{\"u}del}, {Skinner}, {Mel'Nikov},
  {Audard}, {Telleschi}, \& {Briggs}}]{GudelSkinner2007}
{G{\"u}del}, M., {Skinner}, S.~L., {Mel'Nikov}, S.~Y., {Audard}, M.,
  {Telleschi}, A., \& {Briggs}, K.~R. 2007, \aap, 468, 529

\bibitem[{{G{\"u}del} \& {Telleschi}(2007)}]{GudelTelleschi2007}
{G{\"u}del}, M., \& {Telleschi}, A. 2007, \aap, 474, L25

\bibitem[{{Gullbring}(1994)}]{Gullbring1994}
{Gullbring}, E. 1994, \aap, 287, 131

\bibitem[{{G{\"u}nther} {et~al.}(2006){G{\"u}nther}, {Liefke}, {Schmitt},
  {Robrade}, \& {Ness}}]{GuntherLiefke2006}
{G{\"u}nther}, H.~M., {Liefke}, C., {Schmitt}, J.~H.~M.~M., {Robrade}, J., \&
  {Ness}, J. 2006, \aap, 459, L29

\bibitem[{{G{\"u}nther} {et~al.}(2007){G{\"u}nther}, {Schmitt}, {Robrade}, \&
  {Liefke}}]{GuntherSchmitt2007}
{G{\"u}nther}, H.~M., {Schmitt}, J.~H.~M.~M., {Robrade}, J., \& {Liefke}, C.
  2007, \aap, 466, 1111

\bibitem[{{Hartmann} {et~al.}(1994){Hartmann}, {Hewett}, \&
  {Calvet}}]{HartmannHewett1994}
{Hartmann}, L., {Hewett}, R., \& {Calvet}, N. 1994, \apj, 426, 669

\bibitem[{{Herbst} {et~al.}(1994){Herbst}, {Herbst}, {Grossman}, \&
  {Weinstein}}]{HerbstHerbst1994}
{Herbst}, W., {Herbst}, D.~K., {Grossman}, E.~J., \& {Weinstein}, D. 1994, \aj,
  108, 1906

\bibitem[{{Huenemoerder} {et~al.}(2007){Huenemoerder}, {Kastner}, {Testa},
  {Schulz}, \& {Weintraub}}]{HuenemoerderKastner2007}
{Huenemoerder}, D.~P., {Kastner}, J.~H., {Testa}, P., {Schulz}, N.~S., \&
  {Weintraub}, D.~A. 2007, \apj, 671, 592

\bibitem[{{Kashyap} \& {Drake}(2000)}]{KashyapDrake2000}
{Kashyap}, V., \& {Drake}, J.~J. 2000, Bulletin of the Astronomical Society of
  India, 28, 475

\bibitem[{{Kastner} {et~al.}(2002){Kastner}, {Huenemoerder}, {Schulz},
  {Canizares}, \& {Weintraub}}]{KastnerHuenemoerder2002}
{Kastner}, J.~H., {Huenemoerder}, D.~P., {Schulz}, N.~S., {Canizares}, C.~R.,
  \& {Weintraub}, D.~A. 2002, \apj, 567, 434

\bibitem[{{Kastner} {et~al.}(2011){Kastner}, {Sacco}, {Montez}, {Huenemoerder},
  {Shi}, {Alecian}, {Argiroffi}, {Audard}, {Bouvier}, {Damiani}, {Donati},
  {Gregory}, {G{\"u}del}, {Hussain}, {Maggio}, \&
  {Montmerle}}]{KastnerSacco2011}
{Kastner}, J.~H., {Sacco}, G.~G., {Montez}, R., {Huenemoerder}, D.~P., {Shi},
  H., {Alecian}, E., {Argiroffi}, C., {Audard}, M., {Bouvier}, J., {Damiani},
  F., {Donati}, J.-F., {Gregory}, S.~G., {G{\"u}del}, M., {Hussain}, G.~A.~J.,
  {Maggio}, A., \& {Montmerle}, T. 2011, \apjl, 740, L17

\bibitem[{{Kastner} {et~al.}(2008){Kastner}, {Zuckerman}, {Hily-Blant}, \&
  {Forveille}}]{KastnerZuckerman2008}
{Kastner}, J.~H., {Zuckerman}, B., {Hily-Blant}, P., \& {Forveille}, T. 2008,
  \aap, 492, 469

\bibitem[{{K\"onigl}(1991)}]{Konigl1991}
{K\"onigl}, A. 1991, \apjl, 370, L39

\bibitem[{{Kurosawa} {et~al.}(2005){Kurosawa}, {Harries}, \&
  {Symington}}]{KurosawaHarries2005}
{Kurosawa}, R., {Harries}, T.~J., \& {Symington}, N.~H. 2005, \mnras, 358, 671

\bibitem[{{Lamzin}(1999)}]{Lamzin1999}
{Lamzin}, S.~A. 1999, Astronomy Letters, 25, 430

\bibitem[{{Ness} {et~al.}(2004){Ness}, {G{\"u}del}, {Schmitt}, {Audard}, \&
  {Telleschi}}]{NessGudel2004}
{Ness}, J., {G{\"u}del}, M., {Schmitt}, J.~H.~M.~M., {Audard}, M., \&
  {Telleschi}, A. 2004, \aap, 427, 667

\bibitem[{{Petrov} {et~al.}(2001){Petrov}, {Gahm}, {Gameiro}, {Duemmler},
  {Ilyin}, {Laakkonen}, {Lago}, \& {Tuominen}}]{PetrovGahm2001}
{Petrov}, P.~P., {Gahm}, G.~F., {Gameiro}, J.~F., {Duemmler}, R., {Ilyin},
  I.~V., {Laakkonen}, T., {Lago}, M.~T.~V.~T., \& {Tuominen}, I. 2001, \aap,
  369, 993

\bibitem[{{Preibisch} {et~al.}(2005){Preibisch}, {Kim}, {Favata}, {Feigelson},
  {Flaccomio}, {Getman}, {Micela}, {Sciortino}, {Stassun}, {Stelzer}, \&
  {Zinnecker}}]{PreibischKim2005}
{Preibisch}, T., {Kim}, Y., {Favata}, F., {Feigelson}, E.~D., {Flaccomio}, E.,
  {Getman}, K., {Micela}, G., {Sciortino}, S., {Stassun}, K., {Stelzer}, B., \&
  {Zinnecker}, H. 2005, \apjs, 160, 401

\bibitem[{{Robrade} \& {Schmitt}(2007)}]{RobradeSchmitt2007}
{Robrade}, J., \& {Schmitt}, J.~H.~M.~M. 2007, \aap, 473, 229

\bibitem[{{Rodriguez} {et~al.}(2010){Rodriguez}, {Kastner}, {Wilner}, \&
  {Qi}}]{RodriguezKastner2010}
{Rodriguez}, D.~R., {Kastner}, J.~H., {Wilner}, D., \& {Qi}, C. 2010, \apj,
  720, 1684

\bibitem[{{Sacco} {et~al.}(2008){Sacco}, {Argiroffi}, {Orlando}, {Maggio},
  {Peres}, \& {Reale}}]{SaccoArgiroffi2008}
{Sacco}, G.~G., {Argiroffi}, C., {Orlando}, S., {Maggio}, A., {Peres}, G., \&
  {Reale}, F. 2008, \aap, 491, L17

\bibitem[{{Sacco} {et~al.}(2010){Sacco}, {Orlando}, {Argiroffi}, {Maggio},
  {Peres}, {Reale}, \& {Curran}}]{SaccoOrlando2010}
{Sacco}, G.~G., {Orlando}, S., {Argiroffi}, C., {Maggio}, A., {Peres}, G.,
  {Reale}, F., \& {Curran}, R.~L. 2010, \aap, 522, A55

\bibitem[{{Schmitt} {et~al.}(2005){Schmitt}, {Robrade}, {Ness}, {Favata}, \&
  {Stelzer}}]{SchmittRobrade2005}
{Schmitt}, J.~H.~M.~M., {Robrade}, J., {Ness}, J., {Favata}, F., \& {Stelzer},
  B. 2005, \aap, 432, L35

\bibitem[{{Stempels} \& {Gahm}(2004)}]{StempelsGahm2004}
{Stempels}, H.~C., \& {Gahm}, G.~F. 2004, \aap, 421, 1159

\bibitem[{{Testa} {et~al.}(2004){Testa}, {Drake}, \& {Peres}}]{TestaDrake2004}
{Testa}, P., {Drake}, J.~J., \& {Peres}, G. 2004, \apj, 617, 508

\bibitem[{{Testa} {et~al.}(2007){Testa}, {Drake}, {Peres}, \&
  {Huenemoerder}}]{TestaDrake2007}
{Testa}, P., {Drake}, J.~J., {Peres}, G., \& {Huenemoerder}, D.~P. 2007, \apj,
  665, 1349

\bibitem[{{Torres} {et~al.}(2008){Torres}, {Quast}, {Melo}, \&
  {Sterzik}}]{TorresQuast2008}
{Torres}, C.~A.~O., {Quast}, G.~R., {Melo}, C.~H.~F., \& {Sterzik}, M.~F. 2008,
  {Young Nearby Loose Associations}, ed. {Reipurth, B.}, 757

\bibitem[{{Ulrich}(1976)}]{Ulrich1976}
{Ulrich}, R.~K. 1976, \apj, 210, 377

\bibitem[{{Vrba} {et~al.}(1993){Vrba}, {Chugainov}, {Weaver}, \&
  {Stauffer}}]{VrbaChugainov1993}
{Vrba}, F.~J., {Chugainov}, P.~F., {Weaver}, W.~B., \& {Stauffer}, J.~S. 1993,
  \aj, 106, 1608

\bibitem[{{Wilms} {et~al.}(2000){Wilms}, {Allen}, \&
  {McCray}}]{WilmsAllen2000ApJ}
{Wilms}, J., {Allen}, A., \& {McCray}, R. 2000, \apj, 542, 914

\end{thebibliography}


\newpage
